\documentstyle[epsfig]{aprim10}
\input{epsf}

\newif\ifAMStwofonts

\ifoldfss
  \ifCUPmtlplainloaded \else
    \NewTextAlphabet{textbfit} {cmbxti10} {}
    \NewTextAlphabet{textbfss} {cmssbx10} {}
    \NewMathAlphabet{mathbfit} {cmbxti10} {} 
    \NewMathAlphabet{mathbfss} {cmssbx10} {} 
  \fi
  \ifAMStwofonts
    \ifCUPmtlplainloaded \else
      \NewSymbolFont{upmath} {eurm10}
      \NewSymbolFont{AMSa} {msam10}
      \NewMathSymbol{\upi}     {0}{upmath}{19}
      \NewMathSymbol{\umu}     {0}{upmath}{16}
      \NewMathSymbol{\upartial}{0}{upmath}{40}
      \NewMathSymbol{\leqslant}{3}{AMSa}{36}
      \NewMathSymbol{\geqslant}{3}{AMSa}{3E}

    \fi
  \fi
\fi 

\ifnfssone
  \newmathalphabet{\mathit}
  \addtoversion{normal}{\mathit}{cmr}{m}{it}
  \addtoversion{bold}{\mathit}{cmr}{bx}{it}
  \newmathalphabet{\mathbfit} 
  \addtoversion{normal}{\mathbfit}{cmr}{bx}{it}
  \addtoversion{bold}{\mathbfit}{cmr}{bx}{it}
  \newmathalphabet{\mathbfss} 
  \addtoversion{normal}{\mathbfss}{cmss}{bx}{n}
  \addtoversion{bold}{\mathbfss}{cmss}{bx}{n}
  \ifAMStwofonts
    \ifCUPmtlplainloaded \else
      \UseAMStwoboldmath
      \makeatletter
      \new@mathgroup\upmath@group
      \define@mathgroup\mv@normal\upmath@group{eur}{m}{n}
      \define@mathgroup\mv@bold\upmath@group{eur}{b}{n}
      \edef\UPM{\hexnumber\upmath@group}
      \new@mathgroup\amsa@group
      \define@mathgroup\mv@normal\amsa@group{msa}{m}{n}
      \define@mathgroup\mv@bold\amsa@group{msa}{m}{n}
      \edef\AMSa{\hexnumber\amsa@group}
      \makeatother
      \mathchardef\upi="0\UPM19
      \mathchardef\umu="0\UPM16
      \mathchardef\upartial="0\UPM40
      \mathchardef\leqslant="3\AMSa36
      \mathchardef\geqslant="3\AMSa3E
    \fi
  \fi
\fi 

\ifnfsstwo
  \DeclareMathAlphabet{\mathbfit}{OT1}{cmr}{bx}{it}
  \SetMathAlphabet\mathbfit{bold}{OT1}{cmr}{bx}{it}
  \DeclareMathAlphabet{\mathbfss}{OT1}{cmss}{bx}{n}
  \SetMathAlphabet\mathbfss{bold}{OT1}{cmss}{bx}{n}
  \ifAMStwofonts
    \ifCUPmtlplainloaded \else
      \DeclareSymbolFont{UPM}{U}{eur}{m}{n}
      \SetSymbolFont{UPM}{bold}{U}{eur}{b}{n}
      \DeclareSymbolFont{AMSa}{U}{msa}{m}{n}
      \DeclareMathSymbol{\upi}{0}{UPM}{"19}
      \DeclareMathSymbol{\umu}{0}{UPM}{"16}
      \DeclareMathSymbol{\upartial}{0}{UPM}{"40}
      \DeclareMathSymbol{\leqslant}{3}{AMSa}{"36}
      \DeclareMathSymbol{\geqslant}{3}{AMSa}{"3E}
    \fi
  \fi
\fi 

\ifCUPmtlplainloaded \else
  \ifAMStwofonts \else 
    \def\upi{\pi}
    \def\umu{\mu}
    \def\upartial{\partial}
  \fi
\fi

\title[Photometric and Spectroscopic Observations of V1280 Sco]{Photometric and Spectroscopic Observations of V1280 Sco}

\author[Naito et al.]
       {Hiroyuki Naito$^1$,  Sahori Mizoguchi$^2$, Akira Arai$^3$, Masayuki Yamanaka$^3$, 
       \newauthor{Shin-ya Narusawa$^1$, Kozo Sadakane$^2$ and Takashi Iijima$^4$}\\
        $^1$Nishi-Harima Astronomical Observatory, Sayo, Hyogo 679-5313, Japan\\
        $^2$Astronomical Institute, Osaka Kyoiku University, Kashiwara, Osaka 582-8582, Japan\\
        $^3$Hiroshima Astrophysical Science Center, Hiroshima University, Higashi-Hiroshima, Hiroshima 739-8526, Japan\\
        $^4$Astronomical Observatory of Padova, Asiago Section, Osservatorio Astrofisico, 36012 Asiago (Vi), Italy}
\date{}

\pagerange{\pageref{firstpage}--\pageref{lastpage}}
\pubyear{2008}

\begin{document}

\maketitle

\label{firstpage}

\begin{abstract}
Photometries of $B$, $V$, $R{\rm c}$, $I{\rm c}$, $y$, $J$, and $K{\rm s}$ bands and low dispersion optical spectroscopic observations of Nova V1280 Sco, started soon after the outburst, are reported.
We show that V1280 Sco is an Fe~{\sc ii} nova and it is going through the historically slowest spectroscopic evolution.
The rapid decline observed in the early phase was caused by formation of a dust shell. 
We  estimate the abundances of CNO using the absorption lines on a spectrum at pre-maximum, and 
find over-abundances by [C/Fe] $\sim$ 1.4, [N/Fe] $>$ 2.0 and [O/Fe] $\sim$ 1.1.
\end{abstract}


\section{Introduction}

V1280 Sco (Nova Scorpii 2007 No. 1) is a classical nova which was  discovered by Y. Nakamura and Y. Sakurai 
on the same night (Feb. 4, 2007) at 9-th visual magnitude (Yamaoka et al. 2007a). Its early spectra were taken by some 
observers and it was classified as an Fe~{\sc ii} type nova (e.g. Naito \& Narusawa 2007, Munari et al. 2007, 
Kuncarayakti et al. 2008), where the Fe~{\sc ii} type nova spectrum is described in Williams (1992). 
Yamaoka et al. (2007b) reported that the spectrum at the pre-maximum stage looked like an early-type supergiant dominated by absorption lines of hydrogen, iron and other metals. It showed a somewhat slow rise to maximum of about
 3.7 in the $V$ band on Feb. 16 (12 days after the discovery), and became a naked eye nova since V382 Vul and V1494 Aql
 recorded in 1999.  It is also remarkable in that V1280 Sco formed a dust shell in the early phase. 
After maximum, it faded slowly for about 12 days, then it declined rapidly in the visual region
by the obscuration due to dust shell(s), the formation of which was directly detected by Chesneau et al. (2008) 
from near-IR and mid-IR observations by the VLTI at ESO.
Das et al. (2008) presented near-infrared studies of V1280 Sco and suggested that the dust is in  clumpy shells. 


\begin{figure} 
\centerline{\epsfxsize=9cm\epsffile{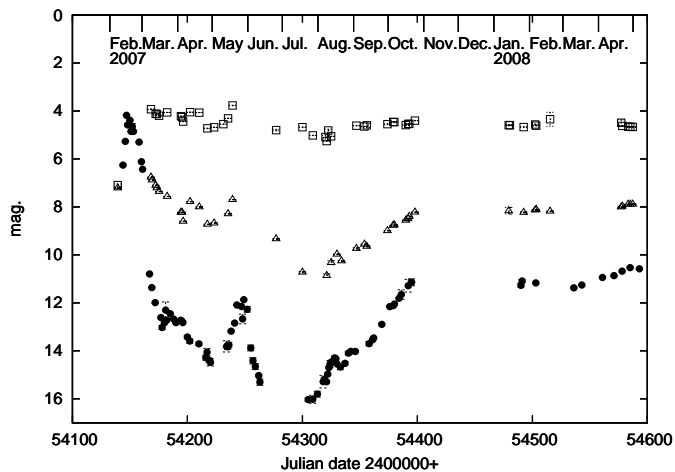}}
\caption[]{Light curves in $V$(filled circle), $J$(open triangle) and $K{\rm s}$(open square) bands.}
\end{figure}

\section{Photometry}
We performed $B$, $V$, $R{\rm c}$, $I{\rm c}$ and $y$ photometries using a 51-cm reflector at Osaka Kyoiku University and 
$J$ and $K{\rm s}$ photometries using the 1.5-m KANATA telescope at Higashi-Hiroshima Observatory. 
The light curves in $V$, $J$ and $K{\rm s}$ bands are shown in Fig. 1. 
Concerning the visual light, within two weeks after maximum, V1280 Sco started a rapid decline until mid-May 
when a re-brightening started. On the other hand, the brightness in the $K{\rm s}$ band stayed at nearly the same 
brightness ($K{\rm s}$ $\sim$ 4) through these days. These results imply that V1280 Sco formed thick dust shell(s), 
which are consistent with the results of Chesneau et al. (2008) and Das et al. (2008). When the first re-brightening
 in the $V$ band occured in the latter half of May 2007, near IR magnitudes had brightened simultaneously. 
We guess that another H-burning event had happened on the WD at that time. The light curves in visual show the 
second re-brightening from the beginning of Aug.,  which is interpreted that the obscuring dust became thinner gradually.

\begin{figure} 
\centerline{\epsfxsize=9cm\epsffile{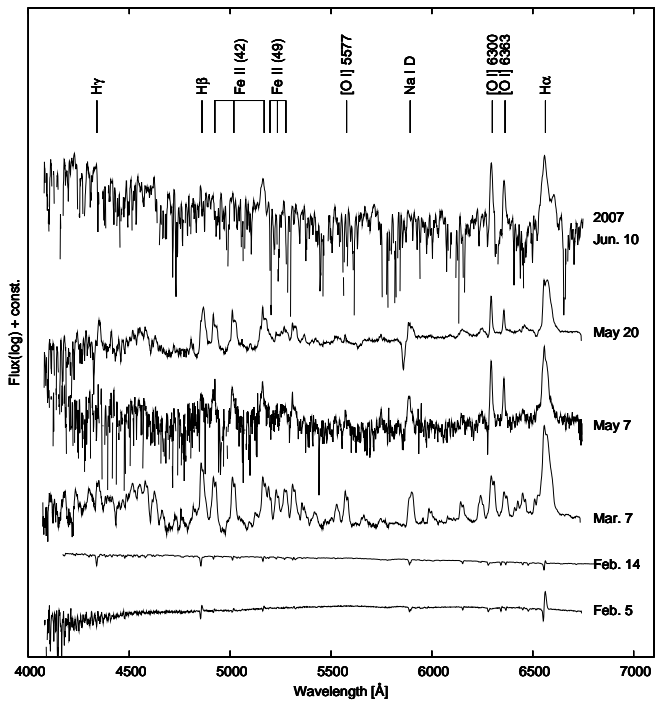}}
\caption[]{Spectral series of V1280 Sco from Feb. 5,  2007 to Jun. 10, 2007. These spectra are corrected for reddening based on the preliminary E(B-V) value of 0.38 determined from photometries near maximum.}
\end{figure}

\begin{figure} 
\centerline{\epsfxsize=9cm\epsffile{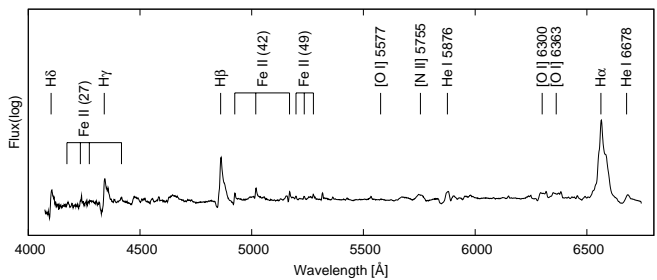}}
\caption[]{The last spectrum taken on Sep. 9, 2008. The reddening is not corrected.}
\end{figure}

\section{Spectroscopy}

We performed optical spectroscopic observations at Nishi-Harima Astronomical Observatory (NHAO), using the 2.0-m 
NAYUTA telescope equipped with the MALLS (Medium And Low dispersion Long slit Spectrograph). We obtained low 
dispersion spectra covering the range 420-680 nm with a resolution of $\lambda$ / $\Delta$ $\lambda$ $\cong$ 1200. 
Fig. 2 shows selected spectra of V1280 Sco observed at NHAO.

The first spectrum taken on Feb. 5 displays a smooth continuum with the Balmer and Fe~{\sc ii} lines in emission, 
accompanied by P Cygni absorptions. The expansion velocity estimated from the blue-shift of the absorption component of H$\alpha$ was 480 km s$^{-1}$. 
This spectrum shows that V1280 Sco 
can be classified as an Fe~{\sc ii} type nova. On Feb. 14, two days before maximum, the spectrum showed a dramatic 
change. It apparently became an early A-type supergiant. From an analysis of absorption lines observed on this spectrum, 
we estimate the abundances of CNO contained of this object, and found over-abundances by [C/Fe] $\sim$1.4, [N/Fe] $>$2.0 
and [O/Fe] $\sim$1.1. The analysis is based on an assumption that the atmospheric parameters of V1280 Sco on Feb. 14 
could be approximated by those of Deneb (A2Ia). 

The spectral evolution after maximum luminosity was very slow and strong forbidden lines of [O~{\sc i}] $\lambda$5577, [O~{\sc i}]  $\lambda$$\lambda$6300,6363 
were observed on Mar. 7. The ratio of [O~{\sc i}] $\lambda$5577 to [O~{\sc i}] $\lambda$ 6300 is so large that we can infer 
the electron density in 
the ejecta was still high because the ejected mass is large and, at the same time, the expansion velocity is low. 
When the optical decline  started by the dust formation from March to mid-May, the continuum flux was too weak to 
be detected and at the same time the H$\beta$ emission also was not detected. However, during the re-brightening phase, H$\beta$ line reappeared. On the same spectrum a P Cygni absorption line having a velocity of 2,000 km s$^{-1}$ at H$\alpha$ was shown. 
These may have been connected with a new mass ejection on the re-brightening. It is caused by the second large outflow 
triggered by a new H-burning event on the WD. Until June 10, no emission line of He~{\sc i}, He~{\sc ii}, or other highly ionized ions such 
as C~{\sc iii}  or N~{\sc iii}  was observed.

On Feb. 18, 2008, the He~{\sc i} $\lambda$6678 appeared and the blue continuum was strong. 
This indicated that the temperature was higher than in 2007, but the temperature 
was not so high as to excite the He~{\sc ii} lines. A higher S/N spectrum observed on Jul. 8 showed P-Cyg profiles 
in the Balmer lines, indicating the gas outflow caused by the continueing H burning. Even on our last spectrum taken 
on Sep. 9, 2008, the forbidden lines of [O~{\sc iii}] $\lambda$$\lambda$4959, 5007 were not seen, thus V1280 Sco has 
not entered the nebular phase yet (Fig. 3).

\section{Summary}
From our observations, it is clear that V1280 Sco is a very unique nova in history. GQ Mus had the H burning 
turn-off time of about 3,000 days (Hachisu et al. 2008), and V723 Cas took 540 days before entering the nebular phase  
(Iijima 2006). V1280 Sco takes longer time than V723 Cas to enter the nebular phase, and it may rewrite the 
record of GQ Mus. Follow-up observations are needed to probe the nature of this very interesting nova.


\label{lastpage}

\clearpage


\begin{thebibliography}{99}
\bibitem[\protect\citename{Chesneau et al. }2008]{chesneau08} Chesneau, O. et al., 2008,
     A\&A, 487, 223
\bibitem[\protect\citename{Das et al. }2008]{das08} Das, R. K. et al., 2008,
     astro-ph/0809.4338
\bibitem[\protect\citename{Hachisu et al.}2008]{hachisu08} Hachisu, I.,  Kato, M., Cassatella, A., 2008,
     astro-ph/0806.4253
\bibitem[\protect\citename{Iijima}2006]{iijima06} Iijima, T., 2006,
     A\&A, 451, 563 
\bibitem[\protect\citename{Kuncarayakti et al. }2008]{kuncarayakti08} Kuncarayakti, H., Kristyowati, D., Kunjaya, C., 2008,
     Ap\&SS, 314, 209
\bibitem[\protect\citename{Munari et al. }2007]{munari07} Munari, U. et al., 2007,
     CBET 852
\bibitem[\protect\citename{Naito \& Narusawa }2007]{naito07} Naito H., Narusawa S., 2007,
     IAC Circ. 8803
\bibitem[\protect\citename{Williams}1992]{williams92} Williams, R. E., 1992,
     AJ, 104, 725
\bibitem[\protect\citename{Yamaoka et al. }2007]{yamaoka07a} Yamaoka, H., et al., 2007a,
     IAU Circ. 8803
\bibitem[\protect\citename{Yamaoka et al. }2007]{yamaoka07b} Yamaoka, H., Fujii, M., Naito, H., 2007b,
     IAU Circ. 8807
\end{thebibliography}
\end{document}